\documentclass[12pt]{article}
\usepackage{graphicx} 

\newcommand{\bq}{\begin{eqnarray}}
\newcommand{\eq}{\end{eqnarray}}

\begin{document}

\title{HJM Local Volatility Model.}
\author{V.M. Belyaev \\
US Bancorp, Minneapolis, MN, USA}


\maketitle
\begin{abstract}
Local Volatility (LV) is a powerful tool for market modeling, enabling the generation of arbitrage-free scenarios calibrated to all European options.

To implement LV, we need to interpolate and extrapolate option prices. This approach is significantly faster and more accurate than any parameterized model. The implementation is demonstrated specifically for interest rate swaptions and caplets. 
A key component of this method is the Small Volatility Approximation within the HJM interest rate model, which is used to calculate sensitivity of forward bond volatility. These calculations are deterministic and fast, with excellent calibration accuracy.

A detailed description of the calibration procedure is provided.
\end{abstract}

\section{Introduction.}

 Interest rates and their dynamics are essential parts of calculations in finance.
 Here we consider SOFR rates \cite{SOFR}  as a risk-free rate.
 
SOFR is a broad measure of the cost of borrowing cash overnight collateralized by Treasury securities. 
The SOFR is calculated directly from transaction data in the US Treasury repurchase market. 
The rate is published by the New York Federal Reserve to serve as a rate upon which other debt transactions can be benchmarked. It was conceived after LIBOR,
 a previously used benchmark rate, suffered a price manipulation scandal in no small part due to its survey based data collection method.
 SOFR rates can be determine from SOFR swaps.
 
 The most difficult part is an interest rate dynamics. There are many models described  in \cite{ALL}. All these models have parameters which supposed to be determine by calibration process.
 It takes time, plus it can not be used to be calibrated to all available swaption prices. The alternative approach is a Local Volatility Model \cite{LV}. It was were developed by Bruno Dupire in 1994.  In the case of equity deterministic
 forward (local) volatility $v_L^2(X,T)$ has the following form:
 \bq
\frac{\partial C(X,T)}{\partial T}+rX\frac{\partial C }{\partial X}=\frac12 v_L^2(X,T)X^2\frac{\partial^2 C(X,T)}{\partial X^2};
\label{LV}
\eq
where $C(X,T)$ represents price of call option with strike $X$ and time to expiration $T$. From (\ref{LV}) we can see that LV model calibration
is just an interpolation and extrapolation of option prices. It can be completed fast and accurate. Notice, that the term $rX\frac{\partial C }{\partial X}$  can be removed if we consider option price in terms of forward equity prices.

It's worth noting that using 
the formula derived by Gatheral \cite{Jim} simplifies the process significantly.
The  Gatheral's formula expresses local volatility in terms of implied as follows:
\bq
v_L^2=\frac{
\frac{dw}{dT}
}
{
1-\frac{y}w\frac{\partial w}{\partial y}+\frac14\left(
-\frac14-\frac1w+\frac{y^2}{w^2}
\right)\left(
\frac{\partial w}{\partial y}
\right)^2+\frac12\frac{\partial^2 w}{\partial y^2}
}.
\label{LV1}
\eq
Here, $w(y,T)$ represent implied variance of an option with strike price $X$ and time to expiration $T$; $y=\ln(X/F(T))$; $F(T)=S_0e^{(r(T)-y(T))T}$ represents the forward price; $S_0$ is the spot price of underlying asset; $r(T)$ and $y(T)$ are interest rate and dividend yield at time $T$, respectively.

Here we will use Normal Volatility Model.
Local Volatility in this case is also known  \cite{LVN}:
\bq
v_L^2=\frac{
\frac{dw}{dT}
}
{
1-\frac{y}w\frac{\partial w}{\partial y}+\frac14\left(
-\frac1{w}+\frac{y^2}{w^2}
\right)\left(
\frac{\partial w}{\partial y}
\right)^2+\frac12\frac{\partial^2 w}{\partial y^2}
};
\label{LVN}
\eq
where  $y=X-F(T)$.

Local Volatility implementation:

\begin{itemize}
\item Equity:\\
Implied Volatilities are determined by Black-Scholes model \cite{BS}. These volatilities can be interpolated and extrapolated to be used in calculation of Local Volatilities.
\item HJM Interest Rate Model \cite{HJM}:\\
Forward Bond implied volatilities  determine model dynamics. 
They can be calculated from swaption prices. Sensitivities of Forward Bond volatilities to the rate shocks can be founded from interpolated and extrapolated swaption prices.
\end{itemize}

 Description of Procedure:

\begin{itemize}
\item Calculate selected swap rate assuming that the bond rate are shifted from ATM rate by some value.
\item This swap rate is used to determine current swaption price (normal volatility, variance) and therefore forward bond volatility.\\
\item Use this procedure to determine rate sensitivities for  all bonds on the grid.
\end{itemize}

Notice, that this approach can be implemented in HJM model  within Small Volatility Approximation. 

To check and to demonstrate accuracy of calculations we use Monte-Carlo simulations with 100,000 scenarios of August 21, 2024 market data.

\section{Small Volatility Approximation.}

In \cite{Belyaev} it is was noticed that Small Volatility Approximation works very well in HJM interest rate models. 

HJM model  is characterized by the following dynamics:
\bq
df(t,T)=\alpha(t,T)dt+\sigma(t,T)dW(t);
\eq
where $f(t,T)$ represent a forward rate:
\bq
B(t,T)=e^{-\int_t^Tf(t,\tau)d\tau};
\eq
$B(t,T)$ denotes a zero coupon risk-free  bond; $\sigma(t,T)$ is a deterministic normal volatility; $dW(t,T)$ represents a Brownian  motion; and
\bq
\alpha(t,T)=\sigma(t,T)\int_t^T\sigma(t,\tau)d\tau;
\eq
is a drift. 

This drift is chosen to satisfy the martingale condition on bond prices
\bq
B(0,T)=\left<
e^{-\int_0^tr(\tau)d\tau}B(t,T)
\right> ;\;\;\forall t\in[0,T].
\eq

In Small Volatility Approximations distribution of discounted bond prices at time $T$ is:
\bq
& & e^{-\int_0^Tr(\tau)d\tau}B(T,T_1)=
\nonumber
\\
& & =B(0,T_1)e^{-\int_0^Td\tau\int_\tau^{T_1}\alpha(\tau,t)dt-\int_0^TdW(\tau)\int_\tau^{T_1}\sigma(\tau,t)dt}=
\nonumber
\\
& & = B(0,T_1)\left(
1-\int_0^TdW(\tau)\int_\tau^{T_1}\sigma(\tau,t)dt+o(\sigma)
\right).
\eq
Within this approximation we can determine  swap price distributions.
Distribution of the SOFR swap  present value   is given by:
\bq
PV(T)=e^{-\int_0^Tr(t)dt}\sum_{n=1}^NB(T,T_n)\left(r_s+1-\frac{B(T,T_{n-1})}{B(T,T_n)}\right)=
\nonumber
\\
=e^{-\int_0^Tr(t)dt}\left(r_s\sum_{n=1}^NB(T,T_n)-
B(T,T)+B(T,T_N)
\right)\simeq
\nonumber
\\
\simeq (r_s-r_{ATM})\sum_{n=1}^NB(0,T_n)+ \Sigma(T,N)\xi\sqrt{T};
\label{PV}
\eq
where $T_n$ represent times to the payments; $r_s$ and $r_{ATM}=\frac{B(0,T)-B(0,T_N)}{\sum_{n=1}^NB(0,T_n)}$ denote  swap  rate  and ATM rate;
 $\xi$ is a standard normal distributed stochastic variable
$$
<\xi>=0;\;\;\;<\xi^2>=1;
$$
\bq
& & \Sigma^2(T,N)T=\int_0^Tv^2(t,N)dt;
\nonumber
\\
& & v(t,N)=r_s\sum_{n=1}^NB(0,T_n)\int_t^{T_n}\sigma(t,\tau)d\tau-
\nonumber
\\
& & -  B(0,T)\int_t^T\sigma(t,\tau)d\tau+B(0,T_N)\int_t^{T_N}\sigma(t,\tau)d\tau.
\label{sva}
\eq

Using formulas (\ref{sva}) we can calculate    swapton prices. It means that we can use them to determine forward bond volatilities from swaption prices.

 To calibrate the model, we have two options:
\begin{itemize}
\item Assume that all unknown volatilities are equal for the selected swaption.
\item Use interpolated volatility surface as input.
\end{itemize}

No significant difference in these two approaches was observed. Here we consider the first way: assuming that all unknown volatilities are equal.

\section{Monte-Carlo Calculations.}

To ensure the quality of calibration, we must use the same input model parameters in both the Small Volatility Approximation and the Monte Carlo simulations. This requires assuming constant volatilities between all time steps. Therefore, the Monte Carlo simulation is implemented as follows:
\bq
f(t_{n+1},T)=f(t_n,T)+\alpha(t_n,T)dt+\sigma(t_n,T)\sqrt{dt}\xi;
\label{df}
\eq
where 
\bq
\alpha(t_n,T)=\frac12  \sigma(t_n,T)^2dt+\sigma(t_n,T)\sum_{k=1}^{n-1}\sigma(t_k,T)dt.
\label{alpha}
\eq
Eq.(\ref{alpha}) can be obtained from martingale conditions:
\bq
B(0,t_n)=B(0,t_k)\left<
B(t_k,t_n)
\right>;\;\; t_n>t_k.
\eq

To calculate swaption prices, we need to find the current SOFR swap price  with yearly payments for the selected scenario as follows:
\bq
& & PV\left(r_X,T_e=Mdt,tenor=(N(N)-M)dt\right)=
\nonumber
\\
& & =e^{-\sum_{m=0}^{M-1}r(t_k)dt}\left(1-e^{-\sum_{k=M}^{N(N)-1}f(t_k,t_{M})dt}-
\right.
\nonumber
\\
& & \left. -r_X\sum_{n=1}^Ne^{-\sum_{k=N(1)}^{N(N)-1}f(t_M,t_k)dt}
\right);
\label{eswp}
\eq

$$
N(k)=M+\frac{k}{dt}.
$$
This implies that the average value of positive and negative values $PV$ provides swaption prices:
\bq
px=\left<[\pm PV(r_X,T_e,tenor)]_+\right>.
\label{swppx}
\eq

Thus, we obtain swaption prices and therefore their implied volatilities using
\bq
px= D(T_e, tenor)\int_{r_X/\left(v\sqrt{T_e}\right)}^\infty \left(v\xi\sqrt{T_e}-r_X\right)e^{-\frac12\xi^2}\frac{d\xi}{\sqrt{2\pi}};
\eq
where $v$ is implied volatility of selected swaption and
\bq
D(T_e,tenor)=\sum_{n=1}^{tenor}e^{-r(T_e+n)(T_e+n)}.
\eq
\section{Calibration Procedure.  }

Consider a grid with 3-month time steps. The first swaption has a tenor of 1 and expires in 3 months. According to equation (\ref{sva}), we have:
\bq
& & v(dt,1)=r_s B(0,5dt)\sum_{k=0}^{4}(k+1)\sigma(0,k ) dt -
\nonumber
\\
& & -B(0,dt)\sigma(0,0)dt+
\nonumber
\\
& & +B(0,5dt)\sum_{k=0}^4(k+1)\sigma(0,k)dt;
\label{1}
\eq
where $dt=0.25$.

Assuming that all unknown volatilities are equal in equation (\ref{1})
\bq
\sigma(0,k)=\sigma(0,0);\;\forall k< 5;
\label{5}
\eq
 we can calculate volatilities in equation (\ref{sva}).

Next, we can extend this calculation to other expirations and tenors, using the previously defined volatility values. For each available tenor and time to expiration  $T_e$ we obtain the following equation:
\bq
D^2(T_e,tenor)\Sigma^2(T_e,tenor)=A\sigma^2+2B\sigma+C;
\label{eq}
\eq
where 
$A, B, C$ are factors which can be determine by using bond  prices and already calculated volatilities; $\sigma$ is an unknown forward volatility. 
In most cases, all parameters are non-negative, so we have
\bq
\sigma=
\frac{1}{A}\left(
-B+
\sqrt{B^2-
A(C-D^2(T_e,tenor)\Sigma^2(T_e,tenor)T_e)
}
\right).
\label{sigma}
\eq

Using this procedure, we can determine all forward bond volatilities. However, it is possible for the discriminant in equation (\ref{sigma}) to become negative. In such cases, we assume the discriminant is zero. The results are shown in Figure 1.

The issue of a potentially negative discriminant can be addressed by more precise modeling of the volatility surface or by employing a different calibration procedure. However, we have observed that the results are only weakly sensitive to these adjustments.
However we need to exclude these negative discriminant problem because they can be a source for instability. The simplest way to do it is to increase parameter $B$ by two times. Then, according to (\ref{eq}) it leads to lower forward volatility. This lower volatility insignificantly changes total swaption volatility but removed most cases of negative discriminant. In many cases it removes all of them. Here we use this
approach for times to expiration more than 2 years.

\section{Forward Bond Sensitivities.}

To generate local volatility (LV) scenarios, we first need to determine forward bond sensitivities. Forward bond volatility depends on the current rate, more precisely on the difference between observed and at-the-money (ATM) rates. This means we must calculate forward volatilities using rates shifted from the ATM level. As a result, the swaption rate will also be adjusted accordingly.

The present value distribution for a swaption with a strike different from ATM, under the Small Volatility Approximation, is given by:
\bq
PV(T_e)\simeq \delta r\sum_{n=1}^{tenor}B(0,T_n)+ \Sigma(\delta r,T_e,tenor)\xi\sqrt{T_e};
\nonumber
\eq
where $T_n=T_e+n$; $\delta r=r_s(T)-r_{ATM}(T)$  is a rate shift..

We can use the same procedure as for ATM swaptions to calculate forward bond volatilities and find the sensitivity of forward bond volatility to bond rates. However, we must account for the fact that the market swap rate differs as follows:
\bq
r_X(T)=\frac{1-e^{-r(T_N)T_N-\delta  rT_N}}{\sum_{n=1}^Ne^{-r(T_n)T_n-\delta  rT_n}};\;\;\; T=T_N;
\nonumber
\eq
and use it in determination of swaption price to find forward bond volatilities:

As we can see from Figures.\ref{p2} and Figures.\ref{m2}  all Monte-Carlo results are in a good agreement with theoretical results (model).
The only exceptions are observed for small time to expirations. It can be explained that we have just a few scenarios for OTM swaptions with strikes different by $2\%$ from ATM strikes.

Thus, we can determine rate volatility sensitivities for every point on our grid.
The simplest way to do it  is to assume that implied variance has a quadratic form for all points on the grid:
	 \bq
 &w(x,t)=\alpha(t)+\beta(t) x+\gamma(t) x^2;
	\label{inter}
	 \eq
\bq
 \alpha(t)=w(0,t);
\;\;\;
 \beta(t)=\frac1{2x_0}(w(x_0,t)-w(-x_0,t));
 \nonumber
 \\
 \gamma(t)=\frac1{x_0^2}(w(x_0,t)-\alpha(t)-\beta(t)x_0);
\eq
where $w(x,t)$ is implied total variance; $x_0=2\%$.

As we can see from Figs.\ref{fig:sm} this assumption does not work very well in case of short-term smiles but
 in case of longer smiles this assumption works good.

Notice, that any linear combination of quadratic smile  has quadratic form. So, we can use this parametrization for forward volatility smiles in case of all expirations.

\section{Swaption Calibration.}

To generate local volatility (LV) scenarios, we calculate the difference between the generated forward rates and the at-the-money (ATM) rates:
\bq
x(T,\tau)= f(T,\tau)-f(0,\tau);
\eq

For a finite time step process, the calculation of the next-step forward rates is given by:
\bq
f(t_{n+1},t_{n+k})=f(t_n,t_{n+k})+\alpha(t_n,t_{n+k})dt+
\nonumber
\\
+\xi v_L(t_{n},t_{n+k})\sqrt{dt};
\label{ten1}
\eq
 where $v_L(t_n,t_{n+k})$ is the deterministic forward bond volatility, as defined by formula (\ref{LVN}) for the normal volatility model.

Using this procedure, we can compute all forward rates and, consequently, the corresponding swaptions. The results for ATM swaptions are shown in Figs.\ref{ATMS}, and the calibration quality appears to be good.

The volatility smiles also look satisfactory, as seen in Figs.\ref{ten1}-\ref{ten30}. However, the 30-year tenor with 30 years to expiration does not exhibit sufficient quality.

\section{Caplets and Floorlets.}

Described procedure can also been used to be calibrated to all caplets and floorlets in addition to swaptions.

Distribution of  present values  of Caplet is
\bq
PV_c(T,r_c)=e^{-\int_0^{T+dt}r(\tau) d\tau}\left(\frac1{dt}\left(\frac{B(T,T)}{B(T,T+dt)}-1\right)-r_c
\right)\simeq
\nonumber
\\
\simeq \frac1{dt}(B(0,T)-(1+r_cdt)B(0,T+dt))+\Sigma_c(r_c,T)\xi\sqrt{T}=
\nonumber
\\
=(r_{ATM}-r_c)B(0,t+dt)+\Sigma_c(r_c,T)\xi\sqrt{T};
\label{cap}
\eq
where  $r_c$ is  caplet rate; $\xi$ is a standard normal distributed stochastic variable;
and
\bq
& & \Sigma^2_c(r_c,T)T=\int_0^Tv_c^2(r_c,t)dt;
\label{cap}
\\
& & v(r_c,t)=B(0,T)\int_t^T\sigma(t,\tau)d\tau-
\nonumber
\\
& & -(1+r_cdt)B(0,T+dt)\int_t^{T+dt}\sigma(t,\tau)d\tau;
\nonumber
\eq
where  caplet market implied volatility is:
\bq
IV(r_c,T)=\frac1{B(0,T+dt)}\Sigma_c(r_c,T).
\eq

We can calculate caplet prices without caplet market and with caplet input.
If we do not use caplets in calibration process then ATM caplet prices are in relative good agreement with market data, but they have significantly lower quality compare to swaption prices Figs.\ref{ATM_Caps}.

To be able generate LV scenarios calibrated to caplets we need to build model for caplet sensitivities. Here we use the same quadratic model as in case of swaptions.
As we can see from Figs.\ref{CapSmiles} interpolation of 6 months smile does not look good, but 1 year smile is more acceptable.

Using these interpolation smiles we can calculate forward bond sensitivities the same way as it was done in case of swaptions.
ATM caplets looks better than without caplet calibration Fig.\ref{AtmCaps}. Short term error can be explained by bad smile interpolation.
The same type of quality we observed for shifted strikes Figs.\ref{CapShifted}. Quality of shifted swaption calibration is the same as in case of
swaption calibration only.
Modeled caplet smiles look also good (see Fig.\ref{cap}).

\section{Conclusions.}

Here, we present the implementation of the HJM Local Volatility Model. 
The results show excellent agreement with input swaption prices across a wide range of expiration dates and tenors. 
Caplet calibration is also in a good agreement with market data. 
This model is suitable for the pricing of all path-independent interest rate derivatives.

\section{Disclaimer.}
The opinions expressed in this article are the author's own and they may be
different from the views of U.S. Bancorp.

\section{Figures}

\begin{figure}[h]
	\begin{minipage}{.5\textwidth}
		\includegraphics[width=50mm]{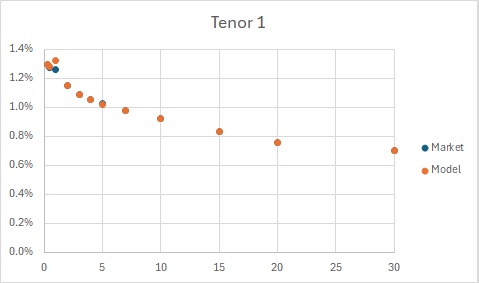}
	\end{minipage}
	\begin{minipage}{.5\textwidth}
		\includegraphics[width=50mm]{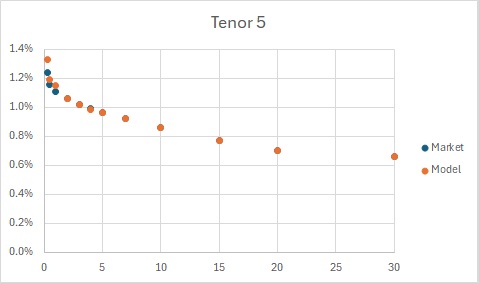}
	\end{minipage}
\end{figure}

\begin{figure}[h]
	\begin{minipage}{.5\textwidth}
		\includegraphics[width=50mm]{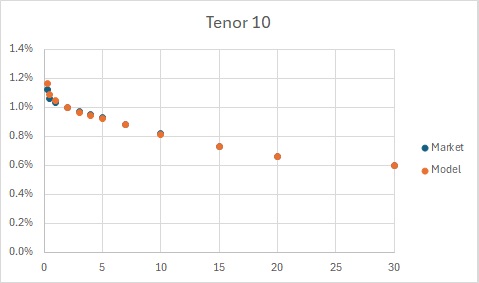}
	\end{minipage}
	\begin{minipage}{.5\textwidth}
		\includegraphics[width=50mm]{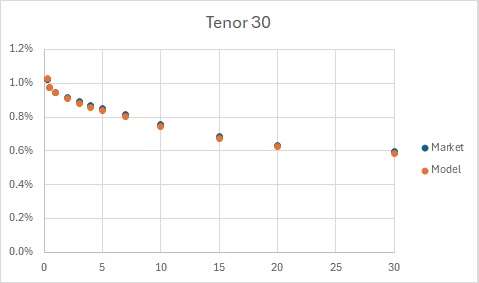}
	\end{minipage}
	\caption{ATM Market and Model Volatilities.}
	\label{ATM}
\end{figure}

\begin{figure}[h]
	\begin{minipage}{.5\textwidth}
		\includegraphics[width=50mm]{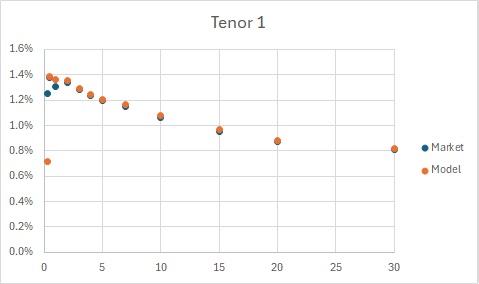}
	\end{minipage}
	\begin{minipage}{.5\textwidth}
		\includegraphics[width=50mm]{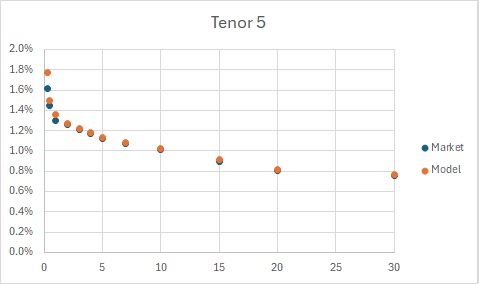}
	\end{minipage}
\end{figure}

\begin{figure}[h]
	\begin{minipage}{.5\textwidth}
		\includegraphics[width=50mm]{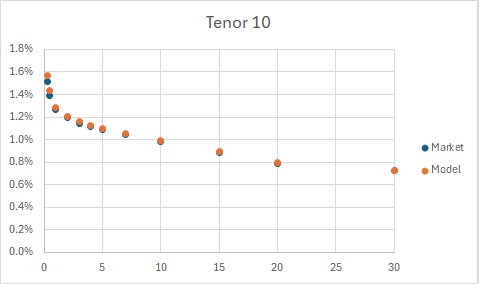}
	\end{minipage}
	\begin{minipage}{.5\textwidth}
		\includegraphics[width=50mm]{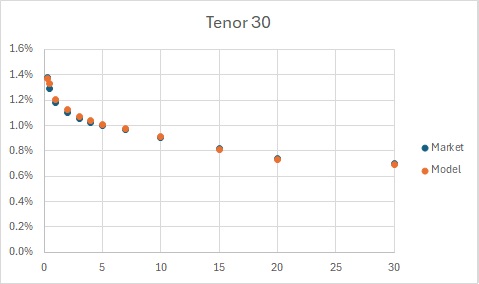}
	\end{minipage}
	\caption{Shifted by 2\% Rate Market and Model Volatilities.}
	\label{p2}
\end{figure}

\begin{figure}[h]
	\begin{minipage}{.5\textwidth}
		\includegraphics[width=50mm]{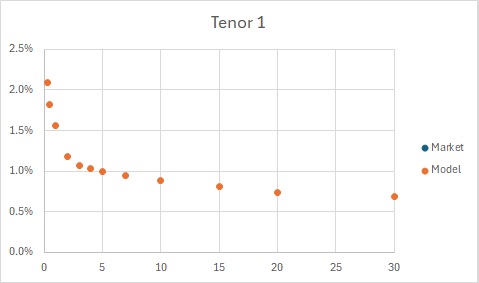}
	\end{minipage}
	\begin{minipage}{.5\textwidth}
		\includegraphics[width=50mm]{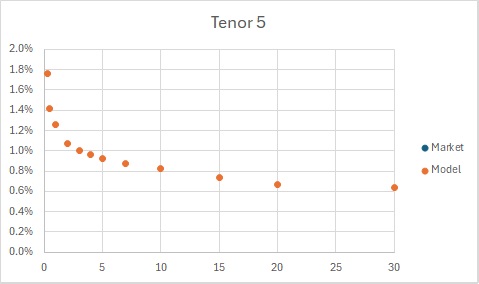}
	\end{minipage}
\end{figure}

\begin{figure}[h]
	\begin{minipage}{.5\textwidth}
		\includegraphics[width=50mm]{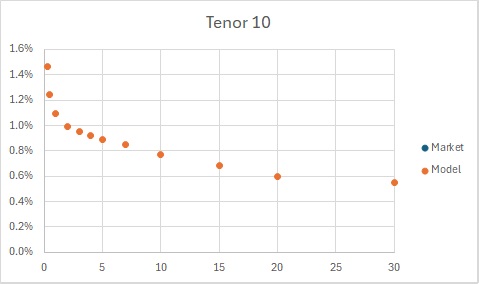}
	\end{minipage}
	\begin{minipage}{.5\textwidth}
		\includegraphics[width=50mm]{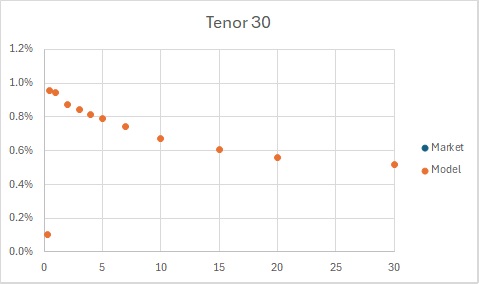}
	\end{minipage}
	\caption{Shifted by -2\% Rate Market and Model Volatilities.}
	\label{m2}
\end{figure}

	\begin{figure}[!h]
		\begin{minipage}{.5\textwidth}
		\includegraphics[width=50mm]{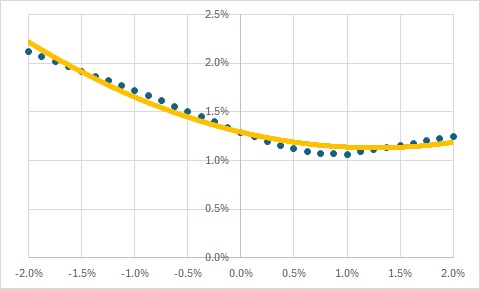}
	\end{minipage}
	\begin{minipage}{.5\textwidth}
		\includegraphics[width=50mm]{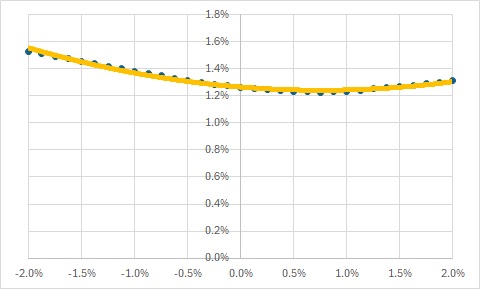}
	\end{minipage}
	\caption{Swaption Volatilities and Quadratic Approximation, Tenor 1.}
		\label{fig:sm}
	\end{figure}

\begin{figure}[h]
	\begin{minipage}{.4\textwidth}
		\includegraphics[width=50mm]{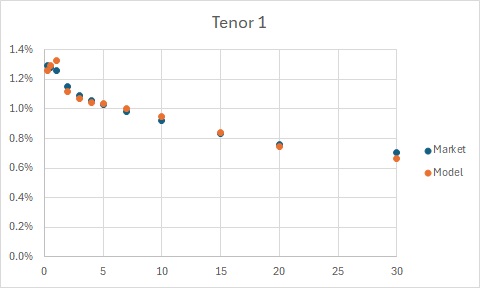}
	\end{minipage}
	\begin{minipage}{.4\textwidth}
		\includegraphics[width=50mm]{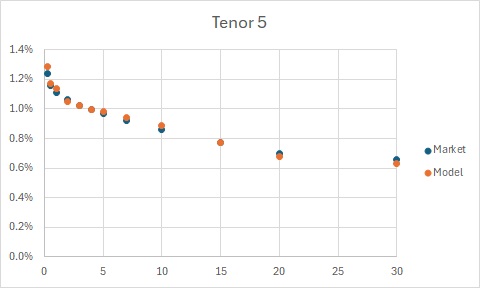}
	\end{minipage}
\end{figure}
\begin{figure}[h]
	\begin{minipage}{.4\textwidth}
		\includegraphics[width=50mm]{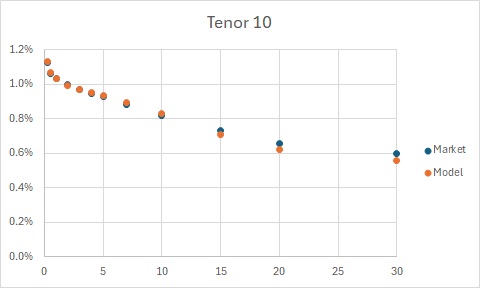}
	\end{minipage}
	\begin{minipage}{.4\textwidth}
		\includegraphics[width=50mm]{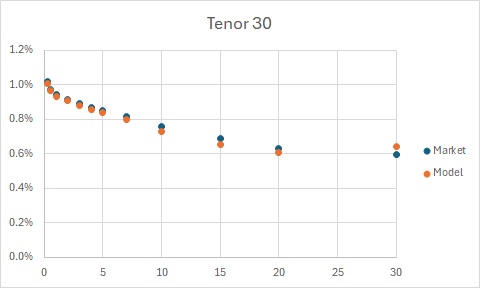}
	\end{minipage}
	\caption{ATM Volatilities. LV Scenarios.}
	\label{ATMS}
\end{figure}
	
\begin{figure}[h]
	\begin{minipage}{.4\textwidth}
		\includegraphics[width=50mm]{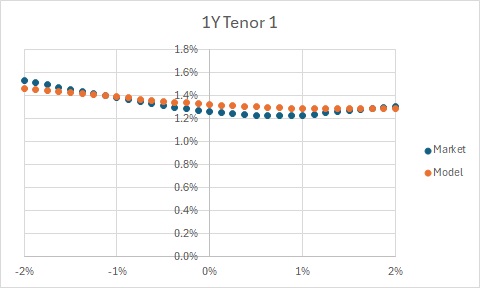}
	\end{minipage}
	\begin{minipage}{.4\textwidth}
		\includegraphics[width=50mm]{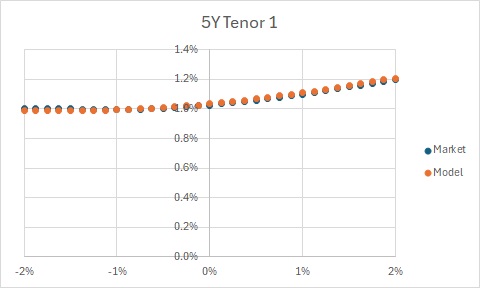}
	\end{minipage}
\end{figure}
\begin{figure}[h]
	\begin{minipage}{.4\textwidth}
		\includegraphics[width=50mm]{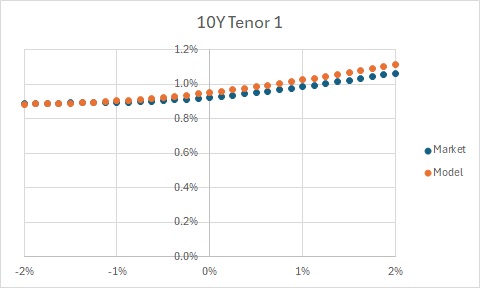}
	\end{minipage}
	\begin{minipage}{.4\textwidth}
		\includegraphics[width=50mm]{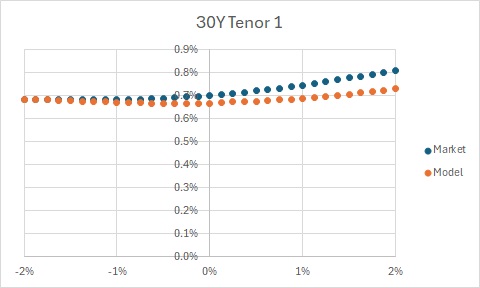}
	\end{minipage}
		\caption{1 Year Tenor Smiles. LV Scenarios.}
	\label{ten1}
\end{figure}
\begin{figure}[h]
	\begin{minipage}{.4\textwidth}
		\includegraphics[width=50mm]{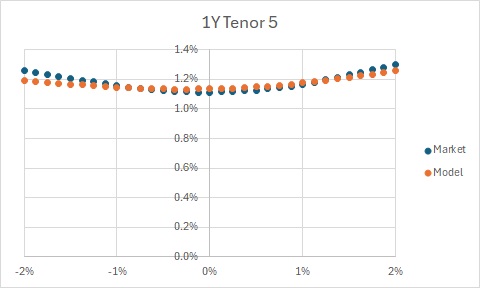}
	\end{minipage}
	\begin{minipage}{.4\textwidth}
		\includegraphics[width=50mm]{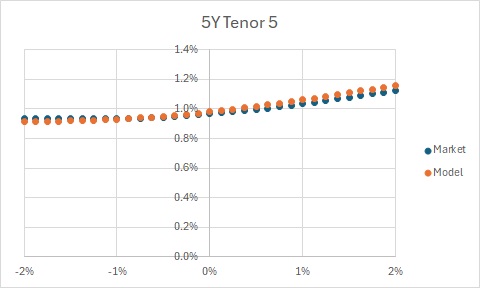}
	\end{minipage}
\end{figure}
\begin{figure}[h]
	\begin{minipage}{.4\textwidth}
		\includegraphics[width=50mm]{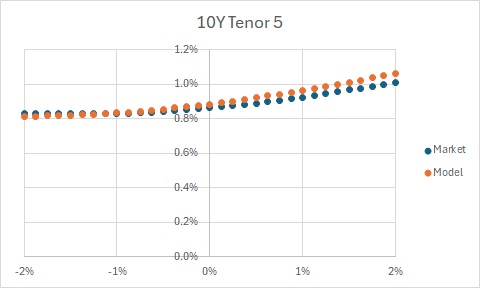}
	\end{minipage}
	\begin{minipage}{.4\textwidth}
		\includegraphics[width=50mm]{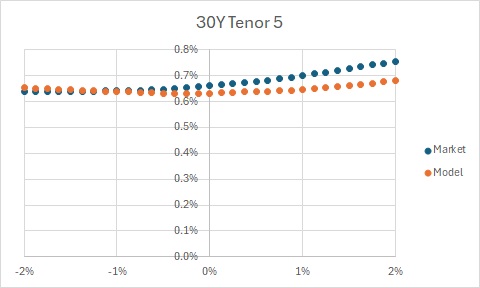}
	\end{minipage}
			\caption{5 Years Tenor Smiles. LV Scenarios.}
	\label{ten5}
\end{figure}
\begin{figure}[h]
	\begin{minipage}{.4\textwidth}
		\includegraphics[width=50mm]{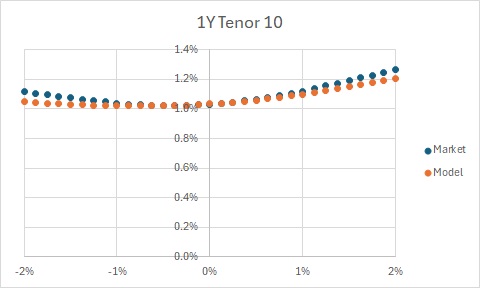}
	\end{minipage}
	\begin{minipage}{.4\textwidth}
		\includegraphics[width=50mm]{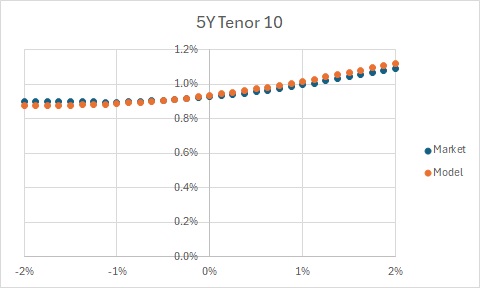}
	\end{minipage}
\end{figure}
\begin{figure}[h]
	\begin{minipage}{.4\textwidth}
		\includegraphics[width=50mm]{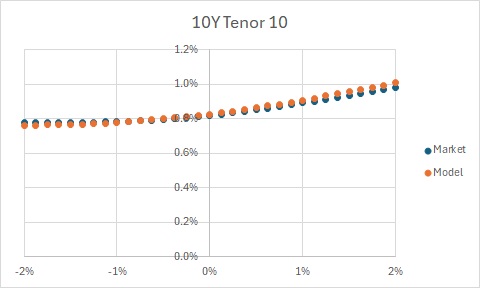}
	\end{minipage}
	\begin{minipage}{.4\textwidth}
		\includegraphics[width=50mm]{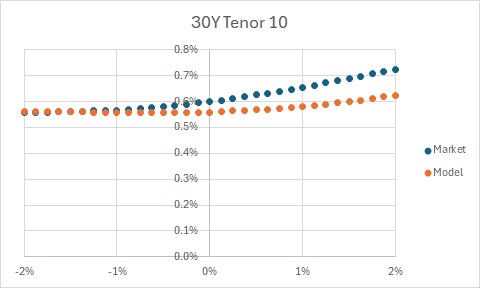}
	\end{minipage}	
			\caption{10 Years Tenor Smiles. LV Scenarios.}
	\label{ten10}
\end{figure}
\begin{figure}[h]
	\begin{minipage}{.4\textwidth}
		\includegraphics[width=50mm]{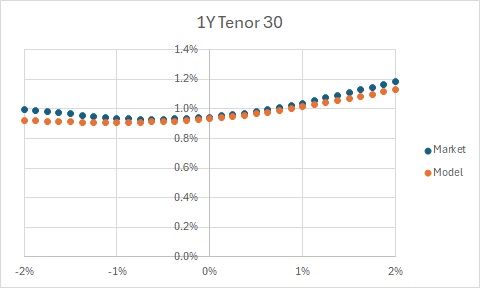}
	\end{minipage}
	\begin{minipage}{.4\textwidth}
		\includegraphics[width=50mm]{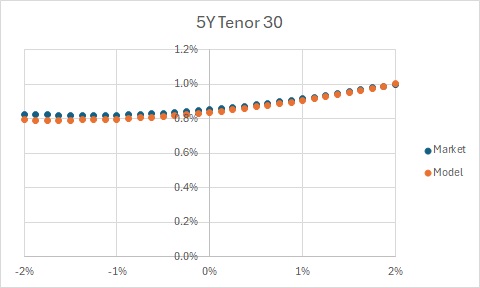}
	\end{minipage}
\end{figure}
\begin{figure}[h]
	\begin{minipage}{.4\textwidth}
		\includegraphics[width=50mm]{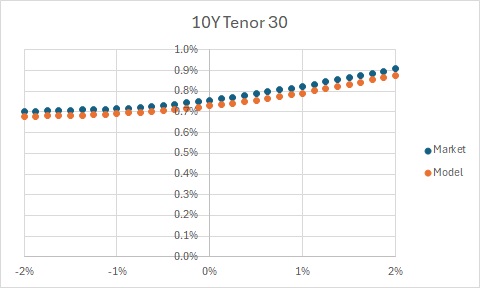}
	\end{minipage}
	\begin{minipage}{.4\textwidth}
		\includegraphics[width=50mm]{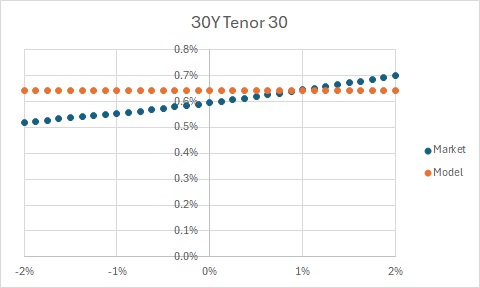}
	\end{minipage}	
			\caption{30 Years Tenor Smiles. LV Scenarios.}
	\label{ten30}
\end{figure}
	
\begin{figure}[h]
	\begin{minipage}{.4\textwidth}
		\includegraphics[width=50mm]{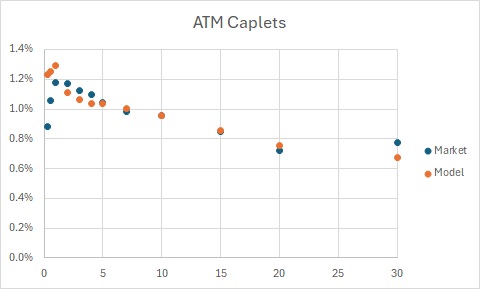}
	\end{minipage}
	\begin{minipage}{.4\textwidth}
		\includegraphics[width=50mm]{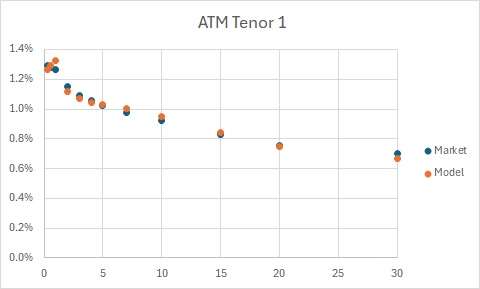}
	\end{minipage}	
		\caption{ATM Volatilities in Interest Rate Model with Calibrated Swaptions only. LV Scenarios.}
			\label{ATM_Caps}
\end{figure}

\begin{figure}[h]
	\begin{minipage}{.4\textwidth}
		\includegraphics[width=50mm]{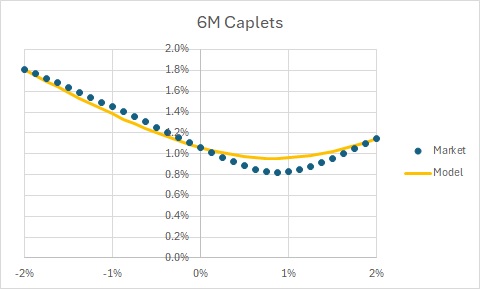}
	\end{minipage}
	\begin{minipage}{.4\textwidth}
		\includegraphics[width=50mm]{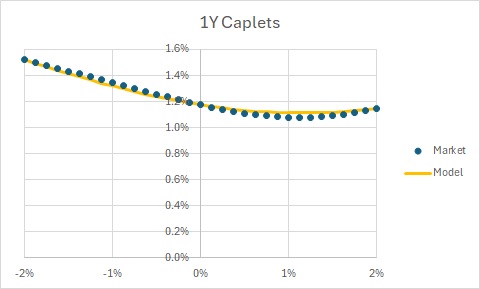}
	\end{minipage}	
		\caption{Caplet Smiles.}
			\label{CapSmiles}
\end{figure}

\begin{figure}[h]
\begin{center}
		\includegraphics[width=50mm]{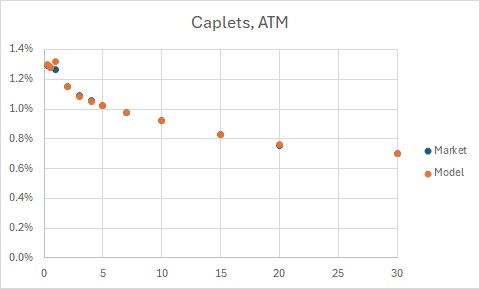}
		\caption{ATM Caplets with Caplet Calibration.}
			\label{AtmCaps}
			\end{center}
\end{figure}

\begin{figure}[h]
	\begin{minipage}{.4\textwidth}
		\includegraphics[width=50mm]{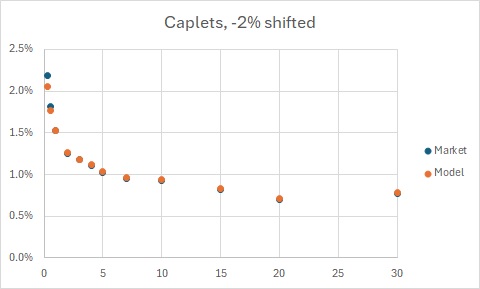}
	\end{minipage}
	\begin{minipage}{.4\textwidth}
		\includegraphics[width=50mm]{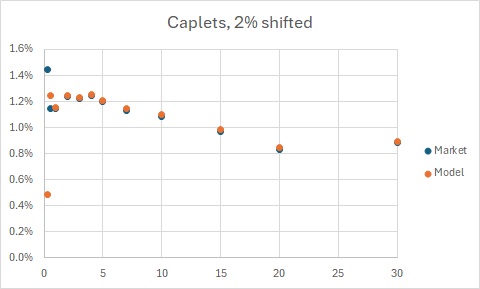}
	\end{minipage}	
		\caption{Shifted Caplet Market and Model Volatilities.}
			\label{CapShifted}
\end{figure}

\begin{figure}[h]
\begin{center}
		\includegraphics[width=50mm]{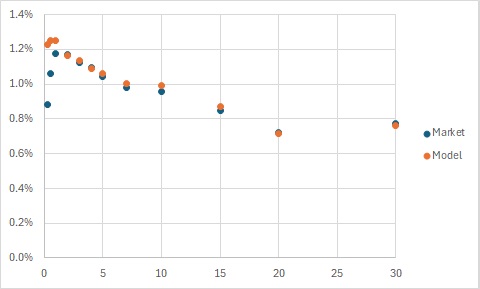}
		\caption{ATM Caplets Volatilities. LV Scenarios.}
			\label{AtmCap}
			\end{center}
\end{figure}

\begin{figure}[h]
	\begin{minipage}{.4\textwidth}
		\includegraphics[width=50mm]{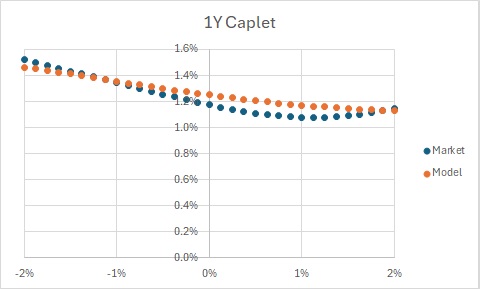}
	\end{minipage}
	\begin{minipage}{.4\textwidth}
		\includegraphics[width=50mm]{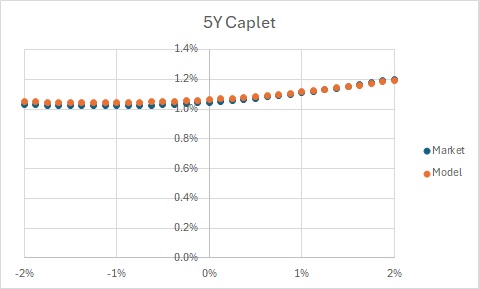}
	\end{minipage}
\end{figure}
\begin{figure}[h]
	\begin{minipage}{.4\textwidth}
		\includegraphics[width=50mm]{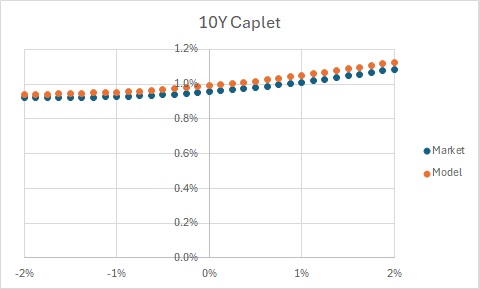}
	\end{minipage}
	\begin{minipage}{.4\textwidth}
		\includegraphics[width=50mm]{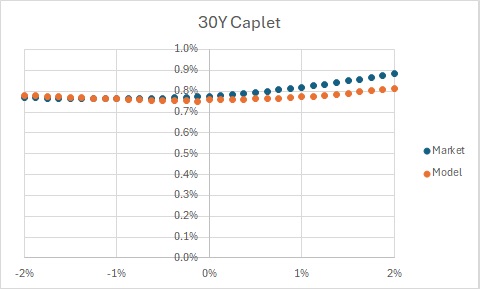}
	\end{minipage}	
		\caption{Caplet Smiles. LV Scenarios.}
			\label{cap}
\end{figure}

\end{document}